 \documentclass[structabstract]{aa}  
\usepackage{graphicx}
\usepackage{subfig}
\usepackage{amssymb}
\usepackage{threeparttable}
\usepackage{natbib}
\bibpunct{(}{)}{;}{a}{}{,} 
\begin{document}
   \title{Hubble Space Telescope study of resolved red giant stars in the outer halos of nearby dwarf starburst galaxies}


   \author{Agnieszka Ry\'{s}
          \inst{1,}
          \inst{2,}
          \inst{3,}
          \inst{4}
          \and
          Aaron J. Grocholski
          \inst{1,}
	  \inst{5}
          \and
          Roeland P. van der Marel
          \inst{1}
          \and
          Alessandra Aloisi
          \inst{1}
		\and
		Francesca Annibali
		\inst{6}
          }

   \institute{Space Telescope Science Institute,
	     3700 San Martin Drive, Baltimore, MD 21218, USA\\
         \and
             Astronomical Observatory, Adam Mickiewicz University,
             S\l oneczna 36, 60-286 Pozna\'{n}, Poland\\
	 \and
	     Insituto de Astrof\'{i}sica de Canarias (IAC), 38200 La Laguna, Tenerife, Spain\\
             \email{arys@iac.es}
	 \and Departamento de Astrof\'{i}sica, Universidad de La Laguna (ULL), 38205 La Laguna, Tenerife, Spain\\
	\and
	     Astronomy Department, Yale University, New Haven, CT 06520, USA\\
	\and
		INAF - Osservatorio Astronomico di Padova, vicolo dell'Osservatorio 5, 35122 Padova, Italy
	}


 
  \abstract
   {Central starbursts in galaxies are an extreme example of ongoing galaxy evolution. The outer parts of galaxies contain a fossil record of galaxy formation and evolution processes in the more distant past. The characterization of resolved stellar populations allows one a detailed study of these topics.}
   {We observed the outer parts of NGC 1569 and NGC 4449, two of the closest
and strongest dwarf starburst galaxies in the local universe, to
characterize their stellar density and populations, and obtain new
insights into the structure, formation, and evolution of starburst
galaxies and galaxy halos.}
   {We obtained HST/WFPC2 images between 5 and 8 scale radii from the
center, along the intermediate and minor axes. We performed point-source photometry to determine color magnitude diagrams of $I$
vs. $V-I$. We compared the results at different radii, including also
our prior HST/ACS results for more centrally located fields.}
   {We detect stars in the RGB and TP-AGB (carbon star) phases in all
outer fields, but not younger stars such as those present at smaller
radii. The RGB star density profile is well fit by either a de Vaucouleurs profile or a power-law profile, but has more stars at large radii than a single exponential. To within the uncertainties, there are no radial gradients in the RGB color or carbon-to-RGB-star ratio at large radii.}
   {The galaxies have faint outer stellar envelopes that are not tidally
truncated within the range of radii addressed by our study. The
density profiles suggest that these are not outward extensions of the
inner disks, but are instead distinct stellar halos. This agrees with
other work on galaxies of similar morphology. 
The presence of such halos is consistent with predictions of hierarchical
galaxy formation scenarios. The halos consist of intermediate-age/old
stars, confirming the results of other studies that have shown the starburst phenomenon
to be very centrally concentrated. There is no evidence for
stellar-population age/metallicity gradients within the halos
themselves.}

   \keywords{galaxies: evolution --
                galaxies: individual (NGC 1569) --
                galaxies: individual (NGC 4449) --
                galaxies: irregular --
		galaxies: photometry --
		galaxies: dwarf --
		galaxies: stellar content
               }
\titlerunning{HST Study of Resolved Red Giant Stars in the Outer Halos of Nearby Starburst Galaxies}
   \maketitle
%

\section{Introduction}

The currently accepted theory of galaxy formation is that galaxies
form hierarchically, with the growth and evolution of large systems
driven by build-up through accretion and merging of smaller units (see
e.g. \citealt{cole:94}). These processes trigger intense, centrally
concentrated bursts of star formation (SF) that provide chemical
enrichment and thermal and mechanical heating of both the interstellar
and intergalactic medium \citep{heckman:98}. While starburst galaxies are thought to be common at high redshift, the
observational information available for such distant galaxies is very
limited owing to their unresolved nature. In order to gain
better understanding of the processes that govern galaxy formation we
turn to nearby starbursting systems. These local analogs of distant starburst galaxies provide the opportunity
for a very detailed study of galaxy formation since we are able to resolve
individual stars in these galaxies with the Hubble Space Telescope (HST)
and thereby build color-magnitude diagrams (CMDs), which are essential
tools for determining ages and abundances of the underlying stellar
populations.

In the local universe, starbursts are typically found in dwarf
irregular galaxies. They are usually gas-rich systems with a chemical
composition comparable to that of distant primeval galaxies
(e.g. \citealt{thuan:99}). Investigating the properties of underlying
older populations in present-day starbursting systems is the key to
understanding their evolution over the cosmological timescale. The
presence of older populations is a proof that a given galaxy has also
experienced episodes of  SF in the past (e.g. \citealt{tosi:91}, \citealt{grebel:98}).

A number of recent studies have aimed at describing the properties and the origin of extended stellar halos that contain these older populations in various galaxy types 
(see e.g. \citealt{seth:07}, \citealt{seth:08}, \citealt{dejong:09} for the results on giant spiral galaxies from the GHOST project, \citealt{tikhonov:06} for the 
massive irregular M82, \citealt{vansevicius:04} for the very low-mass dIrr Leo A). Recent advanced simulations allow one to make predictions regarding 
halo chemical abundance trends \citep{zolotov:10} and their surface brightness profiles following different formation scenarios (e.g. \citealt{cooper:10}); 
they also show that the halos could have formed both as a result of mergers \citep{bekki:08} or because of internal processes in isolated galaxies \citep{stinson:09}.

NGC 1569 and NGC 4449 are two of the closest and strongest starburst
galaxies, at a distance of 3.04 Mpc (Grocholski et al. 2011, in
preparation) and 3.82 Mpc \citep{annibali:08}, respectively. Both have
been studied extensively with HST before (e.g. \citealt{greggio:98},
\citealt{aloisi:01}, \citealt{angeretti:05}, \citealt{blair:98} and
\citealt{gelatt:01}) and are known to display interesting and unusual
properties. NGC 1569 has a star-formation rate (SFR) per unit area that is 2-3 times higher than in other strong starbursts and 2-3 orders of
magnitude higher than in Local Group dwarf irregulars (\citealt{aloisi:01}, \citealt{mcquinn:10}). Its three super star clusters are among the most massive and luminous in the local universe, making the galaxy an extreme example of
a starburst. NGC 4449, with the recent SFR of $\sim$1M$_{\odot}$/yr (\citealt{mcquinn:10}), is the only local ``global'' starburst, in that
star formation appears to occur throughout much of the galaxy
\citep{hunter:97}.

\begin{figure}[!ht]
\centering
\includegraphics[width=0.9\columnwidth]{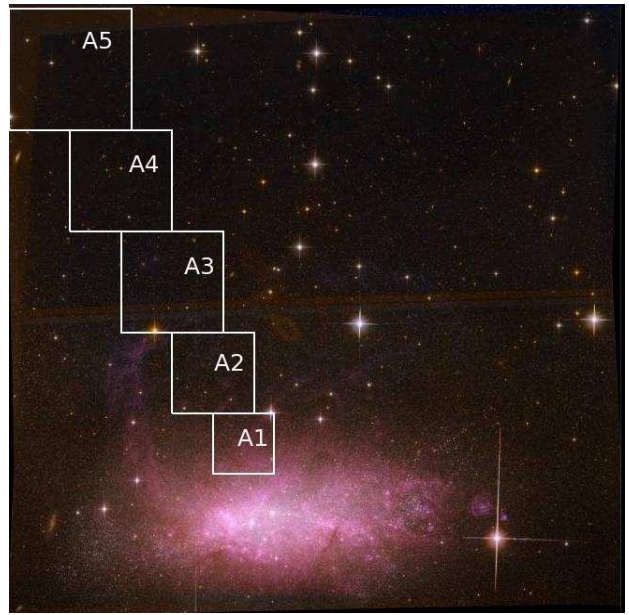}
   \caption{ACS 3-color F658N (H$\alpha$), F814W ($I$) and F606W ($V$) composite image of NGC 1569 showing the locations of ACS subfields. Owing to the decrease in stellar density in the outer portions of the galaxy, we increased the size of our boxes as we move outward from the center of the galaxy.}
\label{1569sub}
\end{figure}

\begin{figure}[!ht]
\centering
\includegraphics[width=0.9\columnwidth]{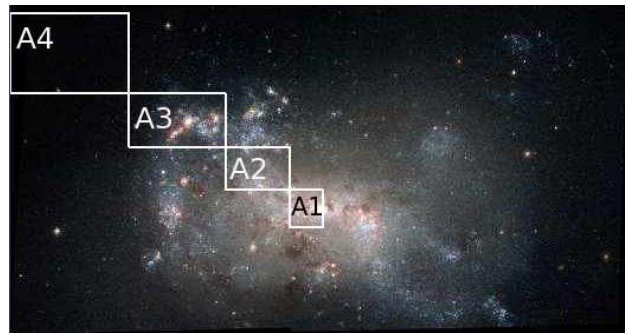}
   \caption{ACS 4-color F435W ($B$), F555W ($V$), F814W ($I$), and F658N (H$\alpha$)
composite image of NGC 4449 showing the locations of ACS subfields. See the caption of Figure~\ref{1569sub} for the explanation of different box sizes.}
\label{4449sub}
\end{figure}

The recent analysis of deep HST/ACS photometry by \cite{grocholski:08}
and \cite{annibali:08} (see Figures \ref{1569sub} and \ref{4449sub}
for mosaicked ACS images) revealed a variety of stellar ages and
metallicities in both galaxies, including well-defined populations of
red giant branch stars. With WFPC2 data that were obtained in parallel
it is possible to extend this analysis to much larger galactocentric
radii, which is the topic of the present paper. We detect and quantify
the presence of old stars at large radii (Section~2). We study stellar
population gradients through variations in the Red Giant Branch (RGB)
color and the relative number of carbon stars as a function of radius
(Section~3). An analysis of density profiles as traced by RGB star
counts allows us to address the question whether the stars at large
radii form a halo component or an extension of the inner disk
(Section~4). Finally, we discuss the results and compare them with the literature (Section~5).

\section{Observations}
\subsection{Fields}
We have observed both NGC 1569 and NGC 4449 with ACS and WFPC2.  In
both cases, ACS was centered on each galaxy with parallel WFPC2
imaging of a region $\sim$~6' away from the galactic centers.

\begin{figure*}
  \centering
\includegraphics[width=0.85\columnwidth]{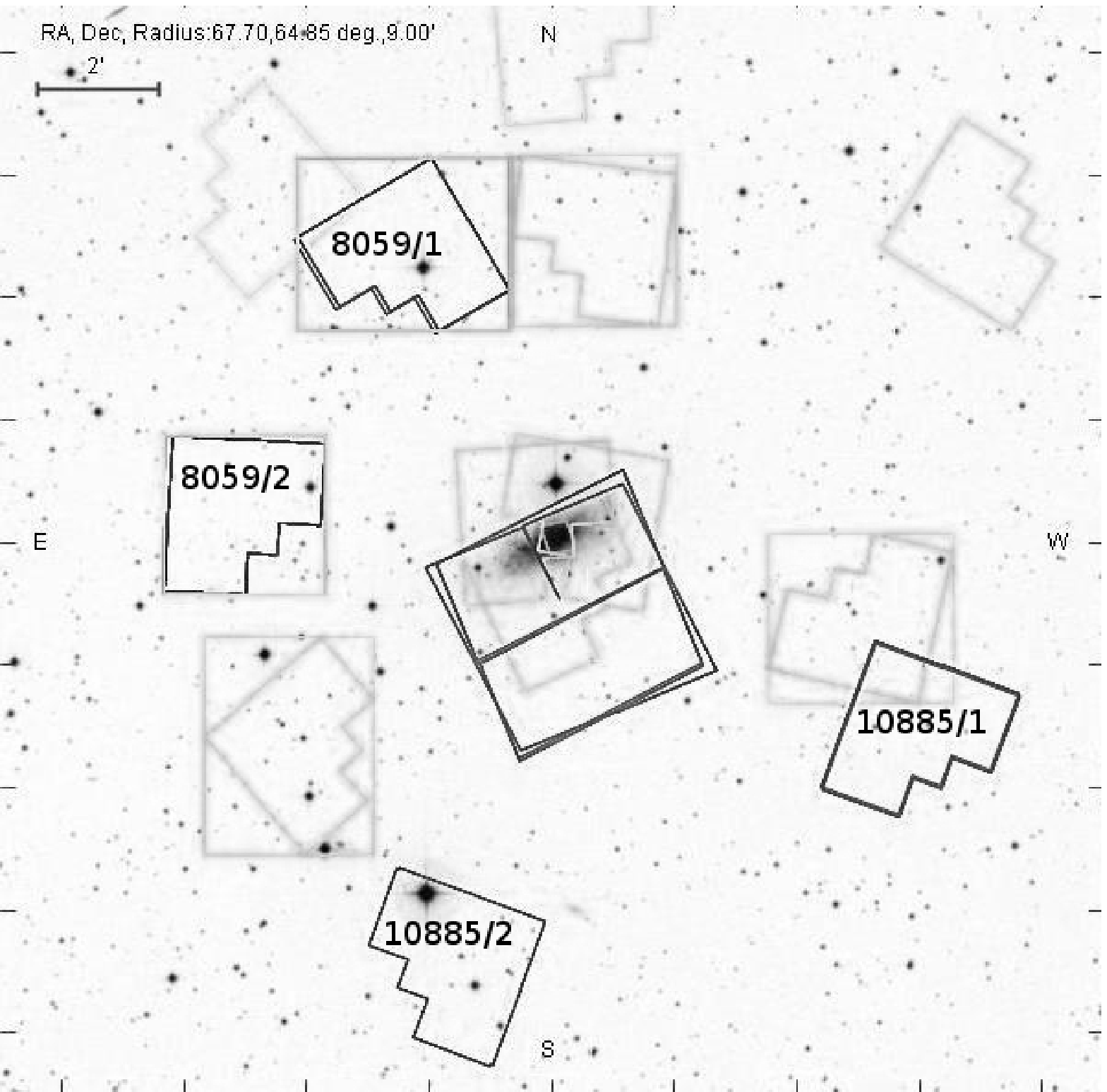}
\includegraphics[width=0.85\columnwidth]{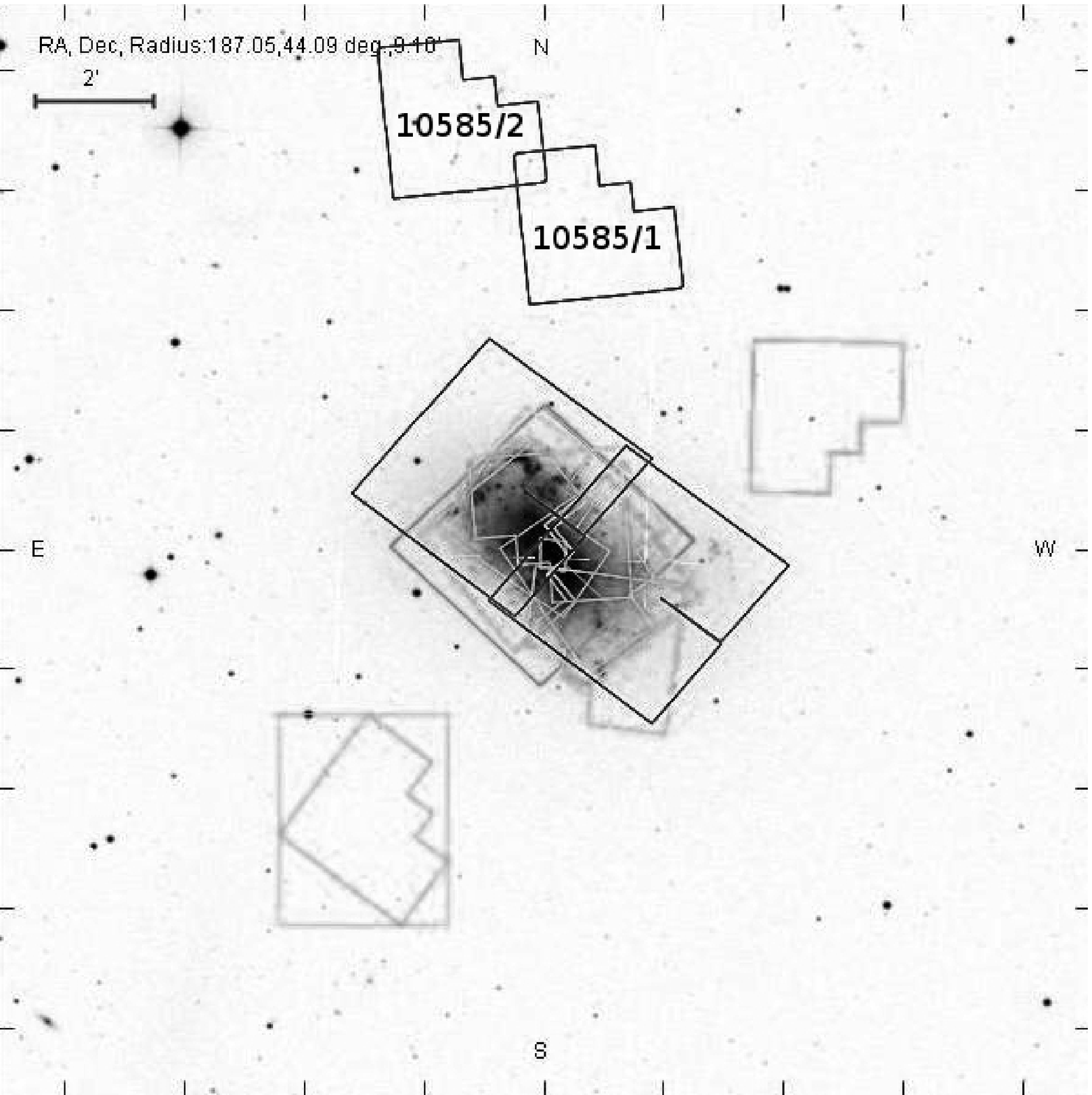}
\caption{Hubble Legacy Archive footprints: ACS and
WFPC2 fields (black) for NGC 1569 (left) and NGC 4449 (right) pointings overlaid on Digitized Sky Survey (DSS) images. The fields depicted in light gray come from programs other than those
listed in Table~\ref{framesinfo} and are not discussed in this paper
because they do not go deep enough to contribute to our analysis.}
\label{hla4449}
\end{figure*}

Our NGC 1569 data were taken during the HST Cycle 15 program 10885 in
November 2006 and February 2007 (PI: A. Aloisi). Between the two sets
of observations, HST was rotated roughly 90 degrees so that, while ACS
was centered on the galaxy, the parallel WFPC2 observations image two
distinct regions in the outskirts of NGC 1569.

NGC 4449 was imaged as a part of the Cycle 14 program 10585 in
November 2005 (PI: A. Aloisi). To fully image the galaxy's center, ACS data were taken at two adjacent,
but overlapping, positions.  As a result, the parallel WFPC2 images are
also adjacent to each other and overlap slightly.

For both galaxies the WFPC2 images were taken in F606W and F814W
broadband filters, which roughly correspond to standard $V$ and $I$
filters in the Johnson-Cousins system. The ACS images were taken in
F606W/F814W for NGC 1569 and in F555W/F814W for NGC 4449. We here present the first photometric analysis of the WFPC2
data and combine this with the previously published ACS data for the
galactic centers (\citealt{grocholski:08}\footnote{For all calculations
in the present paper we use an improved photometric catalog based on a
reanalysis of the data presented in \cite{grocholski:08}.} and
\citealt{annibali:08}).

Table~\ref{framesinfo} lists the coordinates ($\alpha$ and $\delta$),
adopted major axis position angles, axial ratios and distances, number
of frames, total exposure times, galactocentric radii and elliptical
radii (the latter being the major axis radius of the corresponding
isophote). The exact locations of all WFPC2 fields are shown in
Figure~\ref{hla4449} in black.

Other WFPC2 pointings are also available from the HST archive
(programs 6111, 6253, 6423, 7909, 8059, 8133, 8544 9244, 9249 and 9634
for NGC 1569 and 5446, 5971, 6716, 8601, 9244 and 10522 for NGC 4449),
as shown in Figure~\ref{hla4449}. However, only the data from the
GO-8059 program (PI: S. Casertano) are deep enough for our purposes
and are used here.

\begin{table*}
 \begin{threeparttable}[b]
\centering
\caption{Observed fields}
\begin{tabular}{|c|c|c|c|c|c|c|c|c|c|c|}
\hline
\small
Galaxy& $\alpha$ and $\delta$ & \textit{PAmaj} & \textit{q}& Distance & '/kpc scale& Prog./field &Filter& Frames& Exptime& $r$/$m$\\
&(h, $^\mathrm{o}$)&($^\mathrm{o}$)&& (Mpc) &&&&& (s)& (')\\
\hline
NGC    & 4 30 48.8 & 117.0& 0.43 &3.04\tnote{1} &1/0.885 & 10885/1 & F606W & 32 & 33600 &  7.00/14.23 \\
  1569 & +64 50 56.1 && & & & 10885/1 & F814W & 16 & 17600 &  \\
\cline{7-11}
     && &&&&10885/2 & F606W & 24 & 25600 &  7.17/13.33\\
     && &&&&10885/2 & F814W &  4 &  4000 &  \\
\cline{7-11}
     &&&&&&8059/1 & F606W & 3 & 1800 & 5.16/12.03\\
     &&&&&& 8059/1 & F814W & 5 & 3200 &  \\
\cline{7-11}
     &&&&&&8059/2 & F606W & 5 & 5600 & 5.16/7.50\\
     &&&&&&8059/2 & F814W & 6 & 4100 & \\
\hline
NGC  & 12 28 11.2 & 42.4& 0.50& 3.82\tnote{2}& 1/1.111 &10585/1 & F606W & 4 & 3600 &  5.67/8.29\\
4449 & +44 05 36.9 &&&&&10585/1 & F814W & 4 & 6000 &  \\
\cline{7-11}
     &&&&&&10585/2 & F606W & 4 & 3600 &  7.83/10.22\\
     &&&&&&10585/2 & F814W & 4 & 5793 & \\
\hline
\end{tabular}
\label{framesinfo}
\end{threeparttable}
\tablefoot{
Adopted galaxy center coordinates ($\alpha$ and $\delta$),
major axis position angles (\textit{PAmaj}), axial ratios
(\textit{q}), distances and corresponding arcmin/kpc scales; number of
frames, total exposure times and galactocentric radii (\textit{r}) and
elliptical radii (\textit{m}) for each WFPC2 pointing. The axial
ratios and position angles were calculated by taking typical values
from the IRAF \textit{ellipse} task isophote fits at 2 arcmin
radii. Elliptical radii \textit{m} were calculated according to
the following formula: $m^2=x^2 + (y^2/q^2)$ where $x$ and $y$ are
projected major and minor axis distances, respectively.
}
\tablebib{
(1) Grocholski et al. (2011, in preparation); (2)~\citet{annibali:08}.
}
\end{table*}


\subsection{Reduction}
\begin{figure}[!ht]
\centering
\includegraphics[width=0.9\columnwidth]{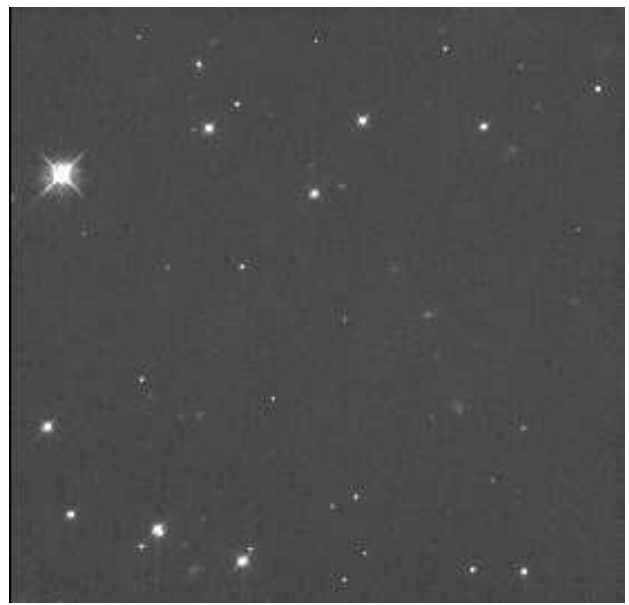}
\caption{Drizzled combination of 32 1000/1100s images for NGC 1569,
field 10885/1, F606W, WF3 chip.}
\label{drizzleresult}
\end{figure}

For each filter and pointing the WFPC2 frames were processed through
the HST data archive on-the-fly reprocessing system, using the most
up-to-date reference files and software. They were then combined into
a single mosaicked image using the software package MultiDrizzle
\citep{koekemoer:02} to improve the sampling of the point-spread
function (PSF) as well as to remove cosmic rays, bad pixels and
correct for geometric distortions. In order to decide on the optimal
\textit{pixfrac} (the linear size of the ``drop'' in input pixels) and
\textit{pixscale} (the size of output pixels) values the so called
``grid experiment'' was carried out, following the recipe given in
\citet{multidrizzle}. Its purpose was to determine what combinations
of the two parameters provide the best resolution and PSF sampling
for our dithered images. The general approach is that the
\textit{pixfrac} should be slightly larger than the \textit{pixscale}
value to allow parts of each ``drop'' to spill over to adjacent
pixels. We found that for most of the images \textit{pixfrac}=0.7 and
\textit{pixscale}=0.6 provided the best resolution and PSF
sampling. An example of a drizzled image is shown in
Figure~\ref{drizzleresult}.

\subsection{PSF photometry}
\begin{figure}[!ht]
\centering
\includegraphics[width=0.9\columnwidth]{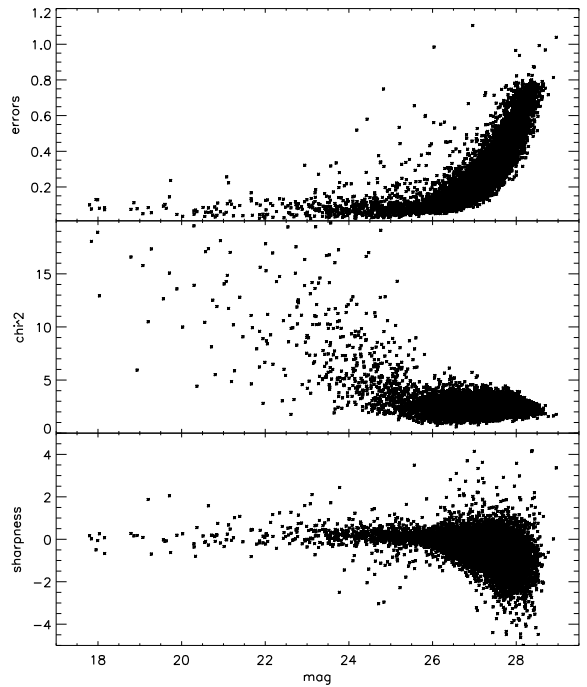}
\caption[Distribution of $\sigma$, sharpness, and
$\chi^2$]{Distribution of the DAOPHOT parameters $\sigma$ (top),
sharpness (middle), and $\chi^2$ (bottom) as a function of F814W
VEGAMAG magnitude (a rough approximation of the Johnson-Cousins I) for
NGC 1569, field 10885/1.}
\label{erchish}
\end{figure}

Point-source photometry was performed with the stand-alone version of
the DAOPHOT/ALLSTAR stellar photometry package \citep{stetson:87}
following the method described in~\citet{grocholski:08}. A set of
plots was created for each pointing showing magnitude versus $\sigma$
(the uncertainty on the magnitude), $\chi^2$ (the residual per degree
of freedom of the PSF-fitting procedure) and \textit{sharpness} (a
measure of the intrinsic size of the object with respect to the PSF
where 0 indicates a ``perfect'' star, a negative value is indicative
of, e.g. bad pixels or cosmic rays, and a positive value indicates
extended objects such as faint background galaxies or unresolved star clusters). An example is
shown in Figure~\ref{erchish} where we can see that the $\sigma$
increases toward fainter magnitudes as expected. The $\chi^2$ values
increase towards brighter magnitudes, which is because
systematic errors in the PSF model are more statistically significant
at lower noise levels. The \textit{sharpness} distribution widens
toward fainter magnitudes, likely indicating increasing contamination
from non-stellar sources.

\subsection{Aperture and CTE corrections}
The PSF magnitudes computed within $\approx$10 pixel radius were corrected to the standard 0\farcs5
aperture for which the filter magnitude zeropoints are known. The
second step was to correct these values to the hypothetical infinite
aperture, which was done by applying a 0.1 mag offset
\citep{holtzman:95}.

Corrections for imperfect charge transfer efficiency (CTE) were
calculated using the prescription by \citet{dolphin:09}. The computed
corrections for most stars were modest, owing to the high sky backgrounds resulting
from long exposures in broad-band filters. However, the corrections
were as high as $\approx$0.14 mag for faint stars located far from the
readout amplifier.

\subsection{Johnson-Cousins transformation}
The conversion of count rates DN/s into the standard Johnson-Cousins
system was performed following the formulae by \citet{holtzman:95} but
using an updated set of zeropoints and transformation coefficients
\citep{dolphin:09}. The transformation formula is of the following
form:
\begin{small}
\begin{equation}SM=-2.5 \cdot \log(DN/s)+T_{1} \cdot SC+T_{2} \cdot SC^2+Z\\ +2.5 \cdot \log(GR),\end{equation}
\end{small}
where SM is the standard magnitude, SC the standard color, T1, T2 the
transformation coefficients, Z the zero point and GR the gain ratio
(slightly different for each chip). It was also
necessary to specify multiple fits for both filters to accurately cover the
entire color range: the transformation formula uses different
coefficients for the color ranges $<$2.0 and $>$2.0.

\subsection{Color-magnitude diagrams}
\begin{figure}[!ht]
  \centering
  \includegraphics[width=0.95\columnwidth]{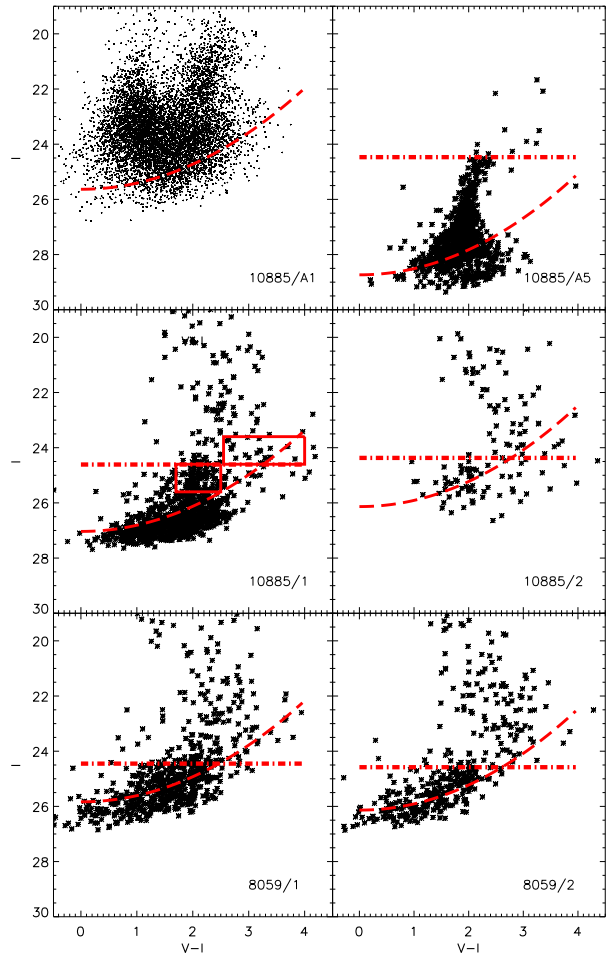}
  \caption{CMDs of the innermost (top left) and outermost ACS (top
  right) subfield and the four WFPC2 fields (middle and bottom) for NGC 1569. The
  dashed-dotted lines indicate the calculated TRGB values and the
  long-dashed lines show 50\% completeness levels. The two boxes shown
  in the 10885/1 plot indicate adopted color and $I$-magnitude ranges
  for selecting RGB and carbon stars (see Sections~\ref{rgbrgb},
  \ref{ccc} and \ref{rrr} for details). Smaller plot symbols are used for the
  A1 subfield to avoid its saturation and make the different
  CMD features visible.}
  \label{1569cmds2}
 \end{figure}

\begin{figure}[!ht]
  \centering
  \includegraphics[width=0.95\columnwidth]{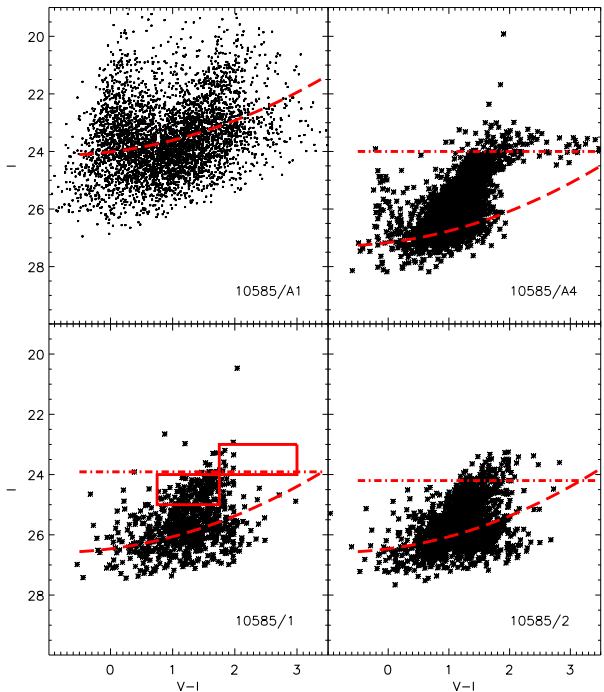}
  \caption{CMDs of the innermost (top left) and outermost ACS (top
  right) subfield and the two WFPC2 fields (bottom) for NGC 4449. See the caption of Figure~\ref{1569cmds2} for further details.}
  \label{4449cmds}
\end{figure}

Figures~\ref{1569cmds2} and \ref{4449cmds} show CMDs of two of our ACS subfields (innermost and
outermost) as well as all of the WFPC2 fields for NGC 1569 and NGC 4449,
respectively. For the WFPC2 fields we used the photometric quality parameters
provided by DAOPHOT to remove false detections. Because the fields are
sparse, fairly lenient cut values were chosen to maintain
clear CMDs.

In the case of NGC 1569 we applied the following cuts to each WFPC2
catalog: $\sigma \le 0.3$ and $-1.0 \le sharpness \le 1.0$. No cuts in
$\chi^2$ were applied since this would reject many real stars at the
bright end of the CMD. A final check was made by visually inspecting
the frames with ``cleaned'' catalogs overplotted to make sure that no
spurious objects were present or no real stars were being
rejected. The catalogs contain 1176, 136, 675 and 391 stars for fields
10885/1, 10885/2, 8059/1 and 8059/2, respectively.

For the NGC 1569 ACS subfields we used the same photometric cuts as in
\cite{grocholski:08}. The subfields contain 7570 (A1), 11173 (A2), 8570
(A3), 3441 (A4) and 926 (A5) stars.

For NGC 4449 there was little evidence for bright stars in the WFPC2
fields. This allowed us to adopt more stringent cut values than for
NGC 1569. Applying the cuts $\sigma\le 0.4$, $-0.5 \le
sharpness \le 0.5$ and $\chi^2<3.0$ allowed us to make sure that we
were neither cutting out real stars nor keeping a lot of false
detections. The final cleaned catalogs contain 827 and 1681 stars in
the first and second field, respectively.

The four NGC 4449 ACS subfields contain 4043 (A1), 11092 (A2), 10897
(A3) and 4510 (A4) stars. Again, the original cuts were applied
\citep{annibali:08}.

\subsection{Completeness}
\label{cmpltnss}
The completeness for all WFPC2 and ACS regions was estimated by
performing artificial star experiments: artificial stars were added to
the frames, each with a randomly assigned magnitude from the range
observed in the real data. Then, the same photometry steps were taken
as for the real data. The completeness was calculated in the individual filters,
and then the final completeness was obtained by multiplying the completeness
factors in the different filters (since in our
photometry we require a star to be detected independently in each
filter). The number of recovered stars divided by the
number of stars used as the input provided the estimate of
completeness as a function of magnitude and position. Both the ACS and WFPC2 50\% completeness limits are shown in Figures
\ref{1569cmds2} and \ref{4449cmds}.

\subsection{ACS photometry}
Here we provide a short summary of the ACS photometric reduction for NGC~1569 and NGC~4449. Specific details can be found in Section~2 of \cite{grocholski:08} and Section~2 of \cite{annibali:08}.

Total exposure times for NGC 1569 were 61716~s in the F606W filter and 24088~s in the F814W filter.  For NGC~4449, the total integration times in F435W, F555W, and F814W were  $\sim$ 3600~s, 2400~s, and 2000~s, respectively.  The individual images were processed through the ACS pipeline (CALACS) and combined into final images using the MULTIDRIZZLE software package \citep{fruchter:09}.

Using the DAOPHOT/ALLSTAR software \citep{stetson:87}, we performed photometry in the following manner.  The PSF template was computed from the most isolated stars in the frame to accurately model any spatial variations in the shape of the PSF.  We then used ALLSTAR to fit the PSF to independent source detection lists for the images in all bands. The photometry lists were then cross-correlated, and the resulting catalogs were cleaned of false detections, background galaxies, or stars with bad measurements by applying both positional and photometric quality cuts. Stars that fell in the gap between the two ACS chips or near the edges of the array were removed because dithering caused these areas to have lower exposure time and therefore more noise.  Detections near bright stars were also removed because diffraction spikes tend to cause false detections. 

We corrected our photometry for CTE losses following the prescription of \cite{riess:04} and then converted instrumental magnitudes into the Johnson-Cousins system following the procedure outlined in \cite{sirianni:05}. Our final catalogs contain 369,213 objects with a measured magnitude in both V and I in the case of NGC 1569 and 402,136 stars in the case of NGC 4449.

%

\section{Stellar population gradients}
\subsection{CMD morphology and foreground contamination}

The innermost ACS CMDs (top left panel in Figures \ref{1569cmds2} and
\ref{4449cmds}) show all the evolutionary features expected at the
magnitudes sampled by our data: a blue plume consisting of MS stars
and blue loop stars at the blue end of their loops, a red plume
consisting of red supergiants and oxygen-rich asymptotic giant branch
(AGB) stars, blue loop stars on tracks that cross between the two
plumes, a horizontal finger of carbon stars extending redward from the
tip of the RGB (TRGB), and a prominent RGB extending from its tip to
the faintest detectable magnitudes. The first three features are due primarily to young stars ($\lesssim
10^{8.5}$ yr), whereas the latter two are primarily due to
intermediate-age and old stars ($\gtrsim 10^{8.5}$ yr).

 \begin{figure}[!ht]
  \centering
  \includegraphics[width=0.95\columnwidth]{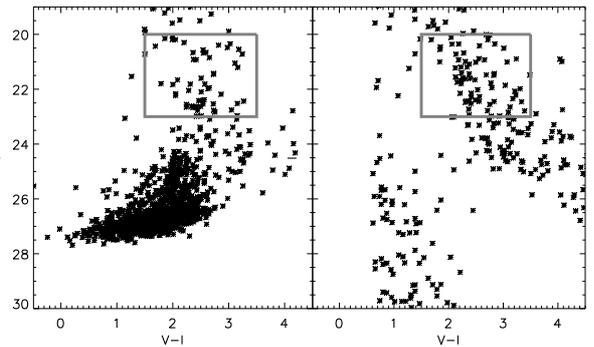}
  \caption{NGC 1569 CMD for field 10885/1 (left) and the Besancon
  model predictions for foreground population for the same field
  (right). The boxes indicate a region populated exclusively by
  foreground stars that was used to calculate the foreground scaling
  factor (see~\ref{ccc}).}
  \label{foreground_example}
\end{figure}

The starbursts in NGC 1569 and NGC 4449 are concentrated toward the galaxy centers. The outermost ACS field of NGC 1569 (10885/A5) shows evidence only for the oldest CMD features of the RGB and carbon stars. These are also the dominant features in the outermost ACS field of NGC 4449 (10585/A4). However, that field still shows a small blue plume due to younger stars at $V-I \approx 0$. This is the outer tail of the younger stars in the main body of NGC 4449, as quantified by the analysis of the spatial dependence of the stellar populations in Annibali et al. (2008; their figures 8 and 19). No evidence of young CMD features is present in any of the large-radii WFPC2 fields that form the primary focus of the present study. We have not attempted to set rigorous limits on any small amounts of young stars at large radii, but no such stars are needed to explain the features evident in the data.

The NGC 1569 WFPC2 CMDs show a red giant branch of stars with colors
(\textit{V-I}) in the $\approx$1.5-2.5 range and a tip at
$I \approx$24.5, a red tail of carbon stars extending above and to the
right of the TRGB (2.4~$\le$~\textit{V-I}~$\le$~4.0,
23.4~$\le$~\textit{I}~$\le$ 24.4) mixed with a strip of foreground
stars located between approximately \textit{V-I}=1.5, \textit{I}=18 and
\textit{V-I}=4, \textit{I}=25. In the case of NGC 4449, the red giant
branch of stars has colors (\textit{V-I}) in the 0.75-1.75 range and a
tip at \textit{I} $\approx$24.0 and the carbon stars are located
approximately in the 1.75~$\le$~\textit{V-I}~$\le$~3.0,
23.0~$\le$~\textit{I}~$\le$ 24.0 region.

NGC 1569 is located close to the Galactic plane ($b=11.24^o$),
therefore our fields suffer from significant foreground
contamination. The contrast of this foreground population with respect
to the stars in the target galaxy increases as one moves further out
in the halo. To quantitatively model the foreground population we used
the Besancon models of stellar population synthesis of the Milky Way
Galaxy \citep{robin:03}. We generated synthetic CMDs for each of the
fields to see how well they match our observational data (see example
in Figure~\ref{foreground_example}). The simulation results fit the
observational CMDs quite well when it comes to the location of
foreground stars on the CMDs but they fail to agree in the exact
number of counts. The predicted number of stars is $\sim$30\% higher
than the number of foreground stars we observe in each field. This is,
however, to be expected as the model is less accurate for low galactic
latitudes and does not account for highly or irregularly absorbing
clouds.

For NGC 4449 ($b=72.40^o$) the Galactic foreground contamination is
negligible. The Besancon model predicts about six stars per square
arcminute down to $I$=28 mag.

\subsection{TRGB magnitude}

The absolute \textit{I}-band magnitude of the TRGB is approximately a
standard candle, with $M_I=-4.0 \pm 0.1$ (\citealt{bellazzini:04},
\citealt{barker:04}). For ages $>$2 Gyr and metallicities ($Z$) $\lesssim$ 0.25 $Z_{\odot}$ there is little variation in the $M_I$ with age or $Z$. Therefore, the observed TRGB magnitude is primarily an indicator of distance, with reddening also playing a role.

Grocholski et al. (2011, in preparation) and \cite{annibali:08} found
the TRGB apparent $I$-band magnitudes for NGC 1569 and NGC 4449 to be
$24.47 \pm 0.04$ and $24.02 \pm 0.05$, respectively. 
Following the approach of the previous authors we calculate the TRGB
magnitude for each of our WFPC2 fields using the method developed by R. P.
van der Marel and described in detail in \citet{cioni:00}.  For NGC
4449, both of our fields excellently agreement with the \cite{annibali:08}
results, with an average of $I_{TRGB} = 24.06 \pm 0.07$.  In NGC 1569,
we find some variation between the TRGB brightness in each of our four
fields, which could be caused by a difference in distance or reddening
amongst the fields.  At the distance of NGC 1569, our outermost field has
a projected distance of 6.35 kpc. If we assume a thin disk geometry
that is inclined 45$^\mathrm{o}$ with respect to the line-of-sight, the difference in
brightness between the TRGB in the outermost field and in the galactic
center is $\sim$0.005 mag. This is well below our photometric
detection limits, therefore suggesting that reddening may be responsible
for differences in the TRGB brightnesses.

The foreground extinction maps of \cite{schlegel:98} have a resolution
of 6.1 arcmin and therefore provide insight into variations in foreground
extinction between our fields.  In general, we find little variation
within each galaxy, with E(V-I) varying by less than $\pm$0.02 mag for three
of the NGC 1569 fields and by less than $\pm$ 0.001 mag for both of the NGC
4449 fields.  This is well within the error bars of the measurements and
can therefore be safely ignored.  The only exception is field 10885/1,
which has an E(V-I) that is 0.15 mag higher than the other NGC 1569
fields.  We therefore correct for this {\it differential} reddening in all
measurements of the 10885/1 field throughout the remainder of this paper to enable a fair comparison to the results of the other fields.

When we correct our TRGB measurement in the 10885/1 field for the
differential reddening, we find the average of all four NGC 1569 fields is
$I_{TRGB} = 24.47 \pm 0.02$, which excellently agrees with previous results.
This agreement provides a useful consistency check on the photometric
calibration, RGB identification, and reddening estimates for our WFPC2
fields.  The individual TRGB determinations are shown as horizontal
dot-dashed lines in Figures~\ref{1569cmds2} and~\ref{4449cmds}.

\subsection{RGB color}
\label{rgbrgb}

\begin{table*}
\centering
\caption{Age and metallicity gradients.} 
\begin{tabular}{|r|r|r|r|r|}
\hline
 & \multicolumn{2}{|c|}{NGC 1569} & \multicolumn{2}{|c|}{NGC 4449} \\ 
\hline
age&\multicolumn{1}{|c|}{3 Gyr}&\multicolumn{1}{|c|}{10 Gyr}&\multicolumn{1}{|c|}{3 Gyr}&\multicolumn{1}{|c|}{10 Gyr} \\
\hline
Z ($\mathrm{\frac{N_{Fe}}{N_{H}}}$)&0.00281$\pm$0.00031&0.00137$\pm$0.00015&0.00370$\pm$0.00030&0.00178$\pm$0.00040 \\
\hline
$\mathrm{Z_{grad}}$(Z/kpc)&0.00007$\pm$0.00007&0.00002$\pm$0.00004&0.00005$\pm$0.00033&0.00002$\pm$0.00015 \\
\hline
$\mathrm{age_{grad}}$(Gyr/kpc)&0.16$\pm$0.19&0.45$\pm$0.64&0.11$\pm$0.45&0.50$\pm$1.50 \\
\hline
\end{tabular}
\label{gradients}
\tablefoot{
Metallicity, metallicity gradient (assuming ages of 3 Gyr and
10 Gyr, respectively) and age gradient (assuming constant $Z$)
estimates for each galaxy, based on the observed RGB color
(gradient).
}
\end{table*}

In order to explore stellar population properties as a function of
distance from the galaxy center we used the WFPC2 fields together with
the ACS subfields shown in Figures~\ref{1569sub} and~\ref{4449sub}.
The WFPC2 fields for NGC 4449 lie on the same side of the galaxy, so
we chose the ACS subfields along the diagonal of the ACS image that
points roughly toward the WFPC2 fields. The WFPC2 fields for NGC 1569
are distributed azimuthally around the galaxy, and we chose the ACS
subfields somewhat arbitrarily to lie along the direction toward the
top left of the ACS image.

\begin{figure}[!ht]
\centering
  \subfloat[NGC 1569]{\label{color1}\includegraphics[width=0.9\columnwidth]{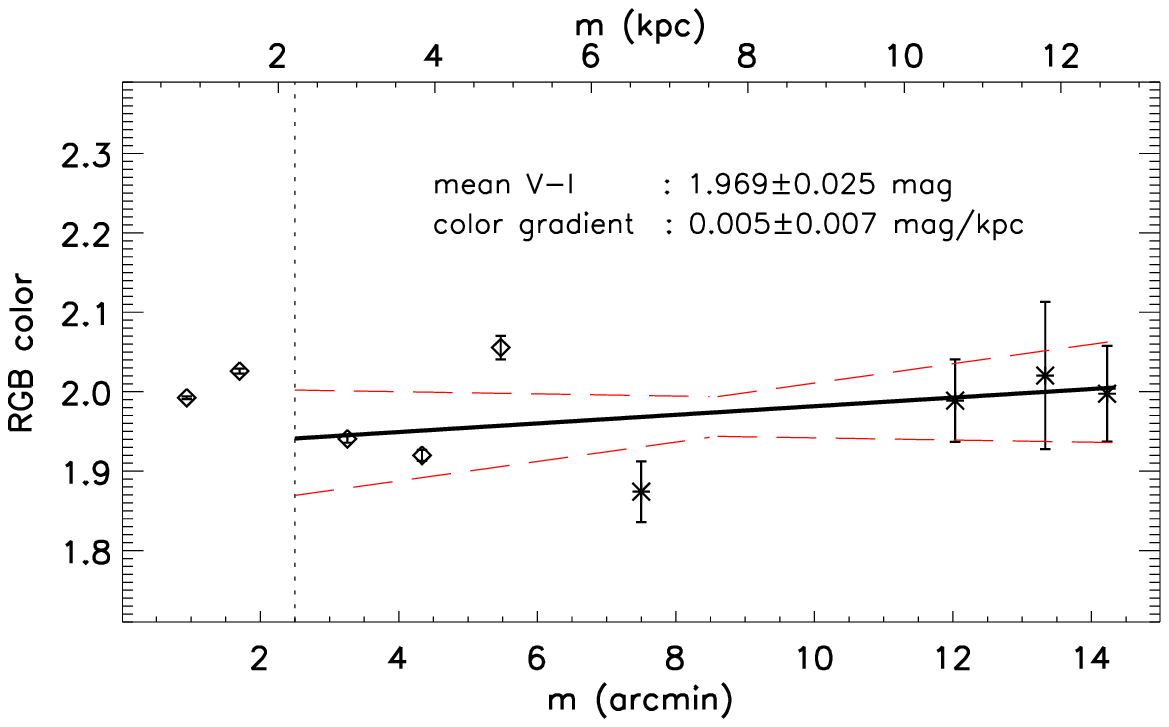}}\\
  \subfloat[NGC 4449]{\label{color2}\includegraphics[width=0.9\columnwidth]{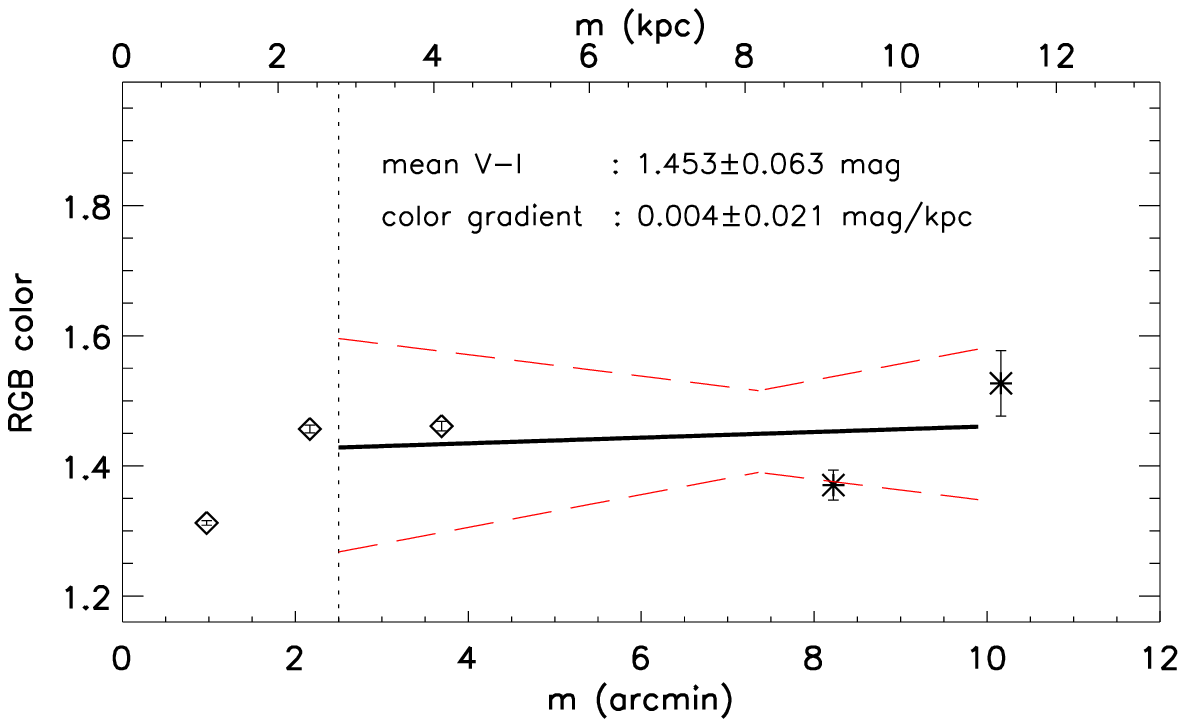}}
   \caption{RGB color variation as a function of elliptical radius $m$
   from the galaxy center. ACS data points are shown as diamonds and
   WFPC2 data points as asterisks. The measured colors shown here are
   not corrected for foreground or (possible) internal
   extinction. Best fits are shown as solid lines and their error
   ranges are indicated by long-dashed lines.}
\label{color}
\end{figure}

To estimate the RGB color, boxes of stars with
1.60~$\le$~\textit{V-I}~$\le$~2.40, 24.4~$\le$~I~$\le$ 25.4 (NGC 1569)
and 0.75~$\le$~\textit{V-I}~$\le$~1.75, 24.0~$\le$~I~$\le$ 25.0 (NGC
4449) were extracted (see Figures~\ref{1569cmds2}
and~\ref{4449cmds}). The stars were then binned according to color and
a Gaussian was fitted to the color distribution. Its peak was taken as
an estimate of the average RGB color and its width divided by the square
root of the number of stars as an estimate of the error. All RGB counts were corrected for the incompleteness
determined as in Section \ref{cmpltnss}. The results are shown as function of elliptical radius
in Figure~\ref{color}.

Near the center of the galaxies the RGB colors are difficult to interpret for two reasons. First, the presence of blue 
loop stars contaminates the blue side of the RGB, which leads to a color that might be biased on the blue side of the true color (as discussed in \citealt{annibali:08}). Second,
there is significant and variable extinction present intrinsic to the target galaxy, as evidenced by patchy dust obscuration (see
e.g. \citealt{kobulnicky:97}). This may lead to an RGB color that is biased on the red side of the true color. Specifically, the central
dip in the NGC 4449 profile is likely caused by the first effect. For
these reasons we do not interpret the RGB colors in the central 2.5
arcmin of each galaxy.

For both galaxies, the color stays relatively constant but with some
up and down variations that exceed the uncertainties in the photometry. 
To quantify this we performed a straight-line fit to the data outside of 
2.5 arcmin. We cast the fit in the form $c=A+B \cdot (m-<m>)$, where $<m>$ 
is the average $m$ of the fields (both in kpc), $c$ is the color, and $A$ 
and $B$ are the average RGB color and color gradient per kpc, respectively. 
There are usually two methods to set the weights in linear fits, one based 
on the known random uncertainties in the data points, and the other based 
on the rms of the points around the best fit. In our analysis we used the 
higher of these two values. This ensures that any unknown systematic 
uncertainties are accounted for in the error bars we derive on A and B. 
The best fits together with their error ranges are shown in Figure~\ref{color}. 
For both galaxies the data are consistent with zero RGB color gradient outside 2.5 arcmin, with the maximum allowed gradients being -0.002 or +0.012 mag/kpc for NGC 1569, and -0.017 or +0.025 mag/kpc for NGC 4449.

The RGB color is a highly degenerate indicator of stellar population
properties, depending on extinction, metallicity, and age. The goal of
the present analysis is, therefore, to search for relative population
gradients and not absolute population properties. Nonetheless, it is
useful to briefly asses the populations that may be consistent with
the observed color. For this we use the stellar population synthesis
model predictions of \cite{girardi:02}. For NGC 1569 we use the
foreground extinction $E(V-I) = 0.60$ from \cite{burstein:84} (see the
discussion in \citealt{grocholski:08}) and for NGC 4449 we use the
foreground extinction $E(V-I) = 0.019$ from \cite{schlegel:98}. We
assume that the internal galaxy extinction is negligible beyond 2.5
arcmin.

Table~\ref{gradients} shows what the measured RGB color implies for
the metallicity $Z$ if one assumes either age = 3 Gyr or 10 Gyr. This
can be compared to the metallicities of the gas in the central region,
which are $Z \approx$0.005 and 0.006 for NGC 1569 and NGC 4449 (see,
e.g. \cite{greggio:98} and \cite{annibali:08} and references
therein). Regardless of the assumed age, the stellar metallicities
found from analyzing the RGB colors are lower than the gas metallicities
in each galaxy. This was to be expected, because the gas has likely been metal-enriched through recent star formation.

Using these results we estimated $Z$ gradients at fixed age and age
gradients at fixed $Z$ (see again Table~\ref{gradients}). The large
error bars mean that the results provide only an upper limit to any
gradient that may be present intrinsically.

\subsection{Carbon star counts}
\label{ccc}

Another CMD feature that allows the study of radial gradients in old
populations is the red tail of carbon stars. A unique identification
of carbon stars requires spectra but as an approximation we can select
stars from the CMD using the region that is believed to be
predominantly composed of carbon stars (see
e.g.~\citealt{marigo:03}).

\begin{figure}[!ht]
\centering
  \subfloat[NGC 1569]{\label{ratio1}\includegraphics[width=0.9\columnwidth]{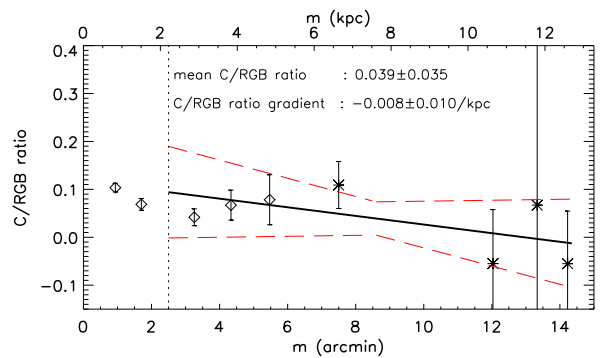}}\\
  \subfloat[NGC 4449]{\label{ratio2}\includegraphics[width=0.9\columnwidth]{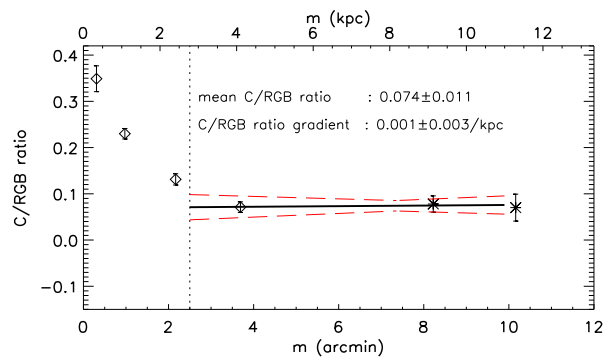}}
   \caption{C/RGB stars ratio as a function of elliptical radius m
   from the galaxy center. ACS data points are shown as diamonds and
   WFPC2 data points as asterisks. Best straight line fits are shown
   as solid lines and their error ranges are indicated by long-dashed
   lines.}
\label{c2rgb}
\end{figure}

To obtain the estimate of the number of carbon stars in NGC 1569 we
first determined a foreground scaling factor by counting stars in a
CMD region that contains only foreground stars (see
Figure~\ref{foreground_example}) in both the data and the Besancon
model and dividing the former by the latter. Then, we subtracted from
the data the foreground model predictions multiplied by this scaling
factor.

Carbon stars were selected at 2.40~$\le$~\textit{V-I}~$\le$~3.85,
23.4~$\le$~$I$~$\le$ 24.4 (NGC 1569) and
1.75~$\le$~\textit{V-I}~$\le$~3.0, 23.0~$\le$~$I$~$\le$ 24.0 (NGC
4449). The color limit was chosen to avoid a significant contribution
of RGB and oxygen-rich AGB stars \citep{annibali:08}. The chosen
ranges are depicted as boxes in the middle left panel of
Figure~\ref{1569cmds2} and the bottom left panel of
Figure~\ref{4449cmds}. Figure~\ref{c2rgb} shows the results as a
function of elliptical radius. The same completeness corrections were applied as for the RGB color. For simplicity, we refer to
the ratio of the counts in the CMD boxes as the C/RGB ratio.  However,
it should be kept in mind that no attempt was made to to uniquely
classify the stars in the chosen boxes.

Using the C/RGB ratio as an absolute indicator of stellar population
properties is difficult for several reasons. For example, the number
of carbon stars depends on the lifetime of the thermally-pulsating (TP)
AGB phase, which is only poorly known. Also, many carbon stars could be
dust enshrouded and unseen in the $I$-band; see e.g. \citet{boyer:09},
who have revealed that up to 40\% of AGB stars may be missing from
optical CMDs. However, the C/RGB ratio can again be used as an
indicator of {\it relative} population gradients. In any stellar
population model the number of TP-AGB and carbon stars tends to
decrease compared to the number of RGB stars with increasing
population age (e.g. \citealt{thuan:05}). Therefore, gradients in
the C/RGB ratio might be indicative of age gradients.

As before, we do not interpret the results inside 2.5 arcmin, because
the ``RGB box'' in the CMD may be contaminated by younger stars that
are not actually on the RGB. Outside of the central 2.5 arcmin the
observed gradients are very shallow. The straight line fits in the
figures show that both galaxies are consistent with zero gradient.

\section{Density profiles}

\begin{figure*}[!ht]
\centering
\includegraphics[width=0.9\columnwidth]{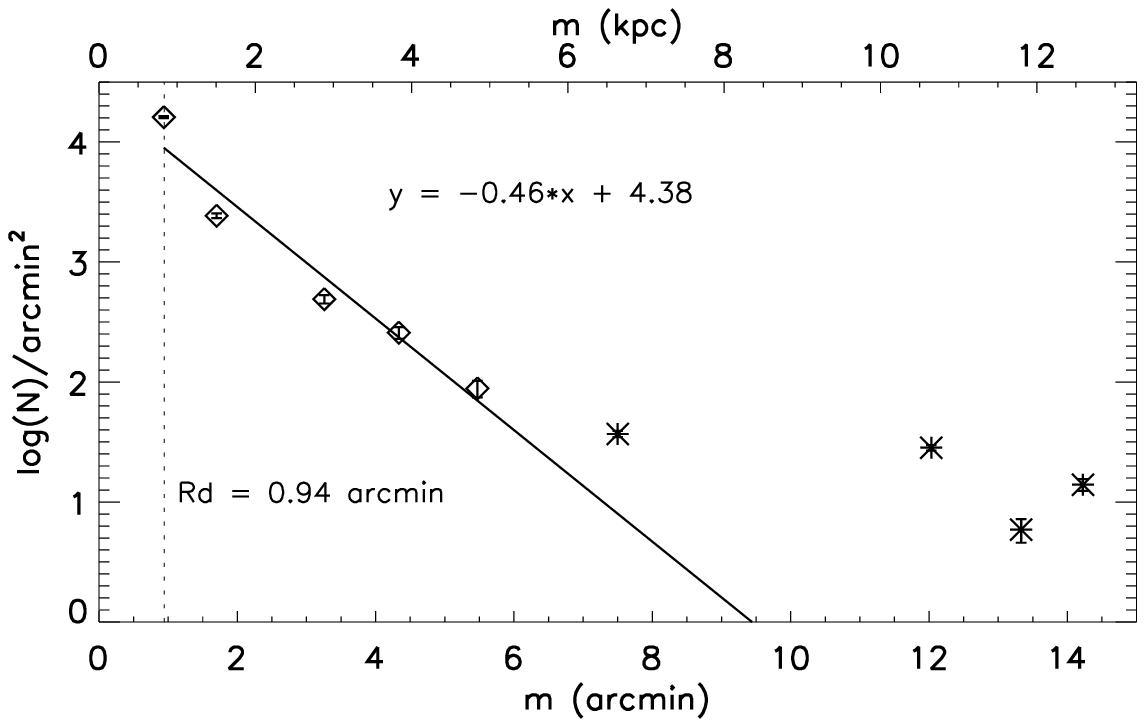}
\includegraphics[width=0.9\columnwidth]{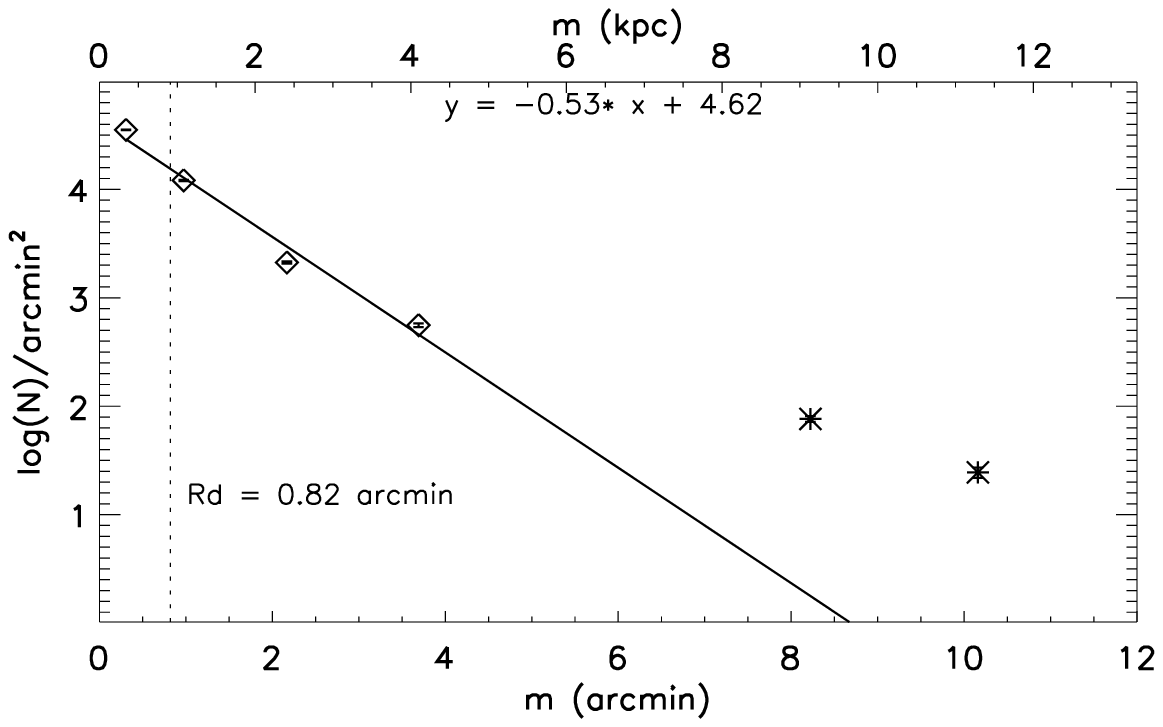}
\includegraphics[width=0.9\columnwidth]{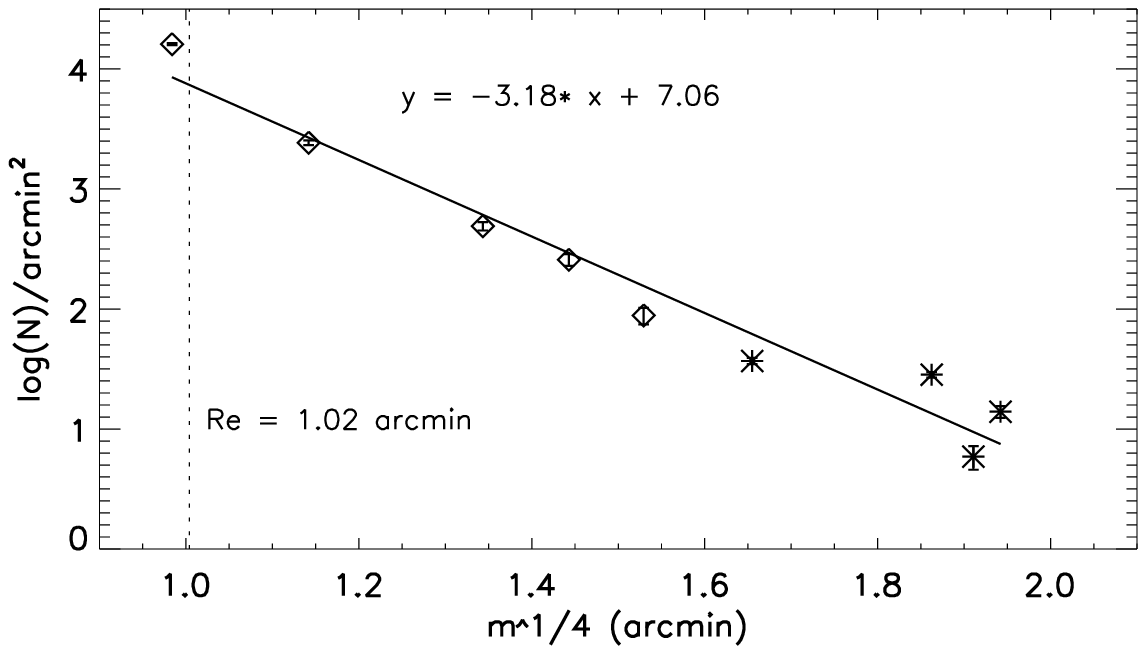}
\includegraphics[width=0.9\columnwidth]{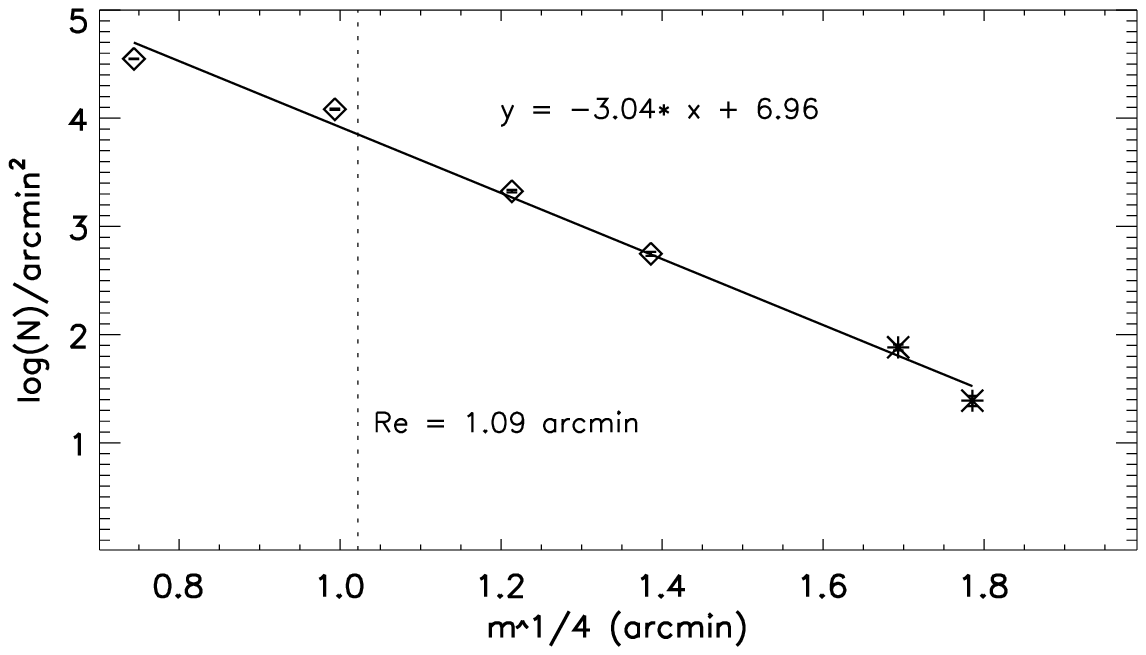}
\includegraphics[width=0.9\columnwidth]{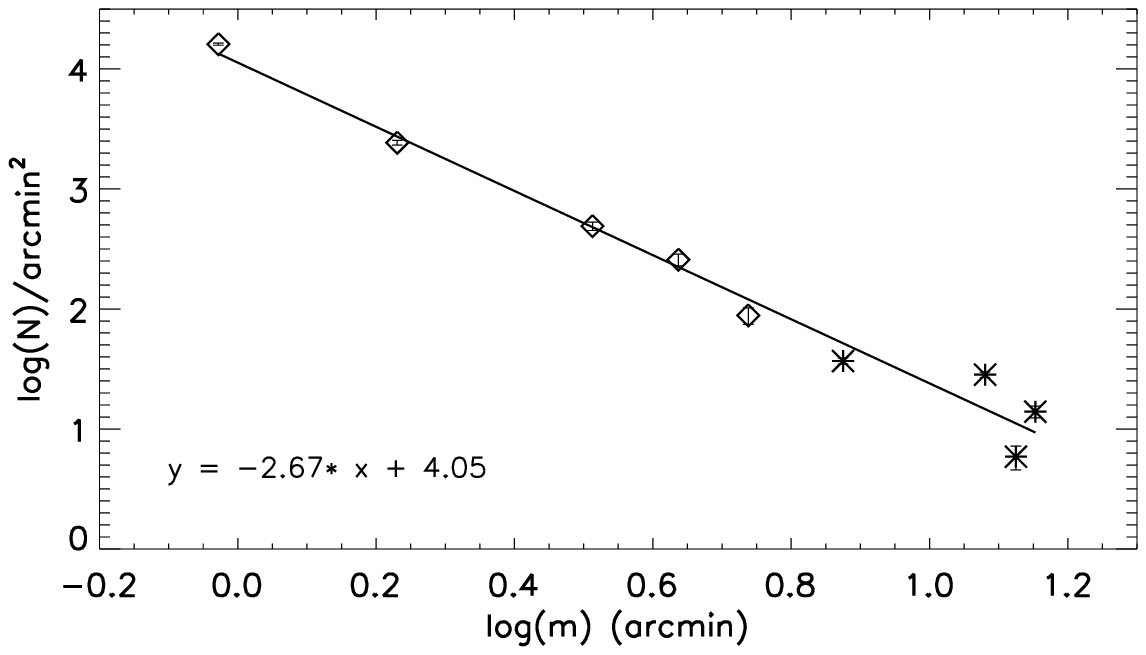}
\includegraphics[width=0.9\columnwidth]{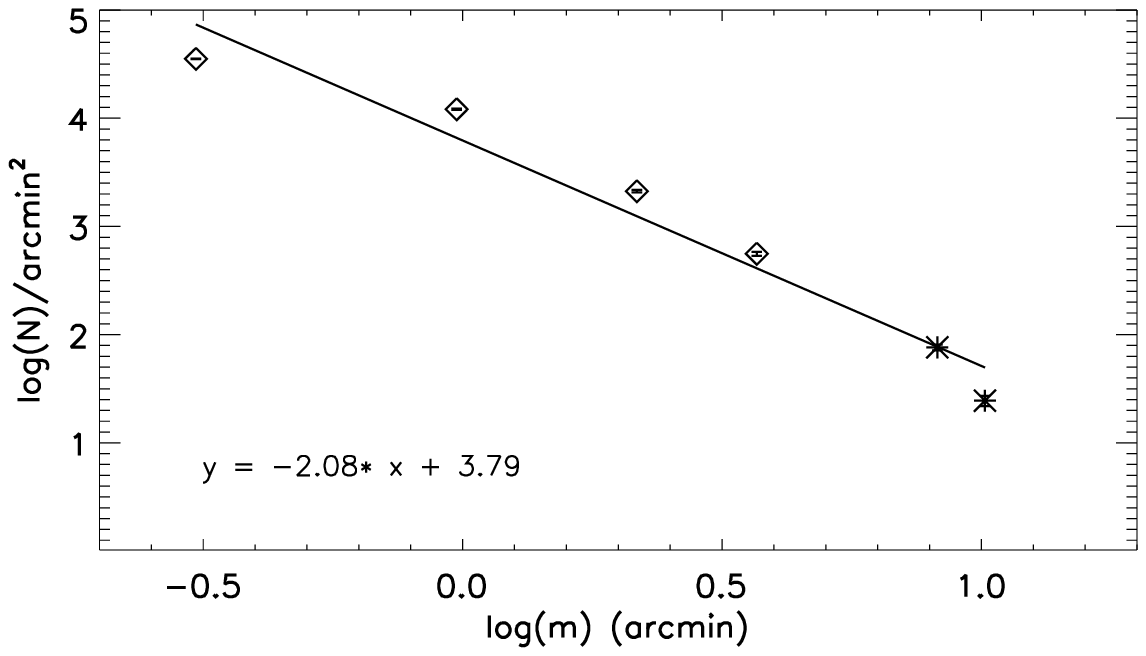}
   \caption[NGC 4449 density profiles]{NGC 1569 (left) and NGC 4449
   (right) density profiles as a function of $m$ (top), $m^{1/4}$
   (middle) and log($m$) (bottom), where $m$ is the elliptical radius
   from the galaxy center. $N$ is the number of stars in the RGB boxes
   shown in Figures~\ref{1569cmds2} and~\ref{4449cmds}. The fits fixed
   to the ACS data points (top) and all points (middle and bottom) are
   shown as solid lines. The symbol types are the same as in
   Figures~\ref{color} and~\ref{c2rgb}. The dotted lines show disk scale lengths $R_d$ (upper panels) and effective radii $R_e$ (middle panels).}
\label{density1}
\end{figure*}

\subsection{RGB counts}
\label{rrr}
Observational evidence indicates that different types of systems
follow different density profiles. For example, the metal-weak halo of
the Milky Way follows a roughly power-law density profile: $\rho \sim
r^{-2.8}$ \citep{juric:08}. Elliptical galaxies tend to have profiles
to large radii that resemble the de Vaucouleurs law: surface
brightness is linear with \textit{r}$^{1/4}$ \citep{schade:99}. Spiral
galaxy disks tend to have an exponential fall-off with radius: their
surface brightness is linear with \textit{r} (and they often have a
truncation at some radius; e.g. \citealt{dejong:07}). These different
profiles are straight lines in plots of log density versus either
$log(m)$, $m^{1/4}$ or $m$, where $m$ is the elliptical radius.

Because we are interested in understanding to what kind of population our
stars belong, we study their density profiles using the number density of
RGB stars at different radii.  To do this, we use the same boxes and
completeness corrections as we did to study the RGB color.  In Figure~\ref{density1} we
plot the density profiles of NGC 1569 (left) and NGC 4449 (right) as a
function of the elliptical radius, $m$ (top), as well as $m^{1/4}$
(middle) and log(m) (bottom).  The equation for the best straight line
fits are given, with the fit overplotted on the data.  For the exponential
fits, the disk scale lengths, $R_d$, are given and $R_e$, the effective
radii, are indicated for the $m^{1/4}$ fits. 

The values of $R_d$ quoted in the figures are higher than the ones previously given in the literature (0.385' for NGC 1569 \citep{willett:05} and 0.763' for NGC 4449 \citep{hunter:05}, both converted from values in kpc using the distances adopted by the authors). This is presumably because previous authors measured integrated light in the optical, where one is sensitive to young stars that are more concentrated toward the galaxy center than the RGB stars studied here.

For both galaxies the star counts appear to be fitted reasonably well
with the de Vaucouleurs and power-law profiles (middle and
bottom panel of Figure~\ref{density1}, respectively). The log($N$) vs. $m$
plots (Figure~\ref{density1}, top panel) show that the number of stars at large
galactocentric distances is higher than a simple exponential model
would predict. This suggests that the stars we see at large radii are
not just the extrapolation of an inner exponential disk. Instead, they
are probably part of a spheroidal stellar halo component.

\begin{figure*}[h]
\centering
\includegraphics[width=0.9\columnwidth]{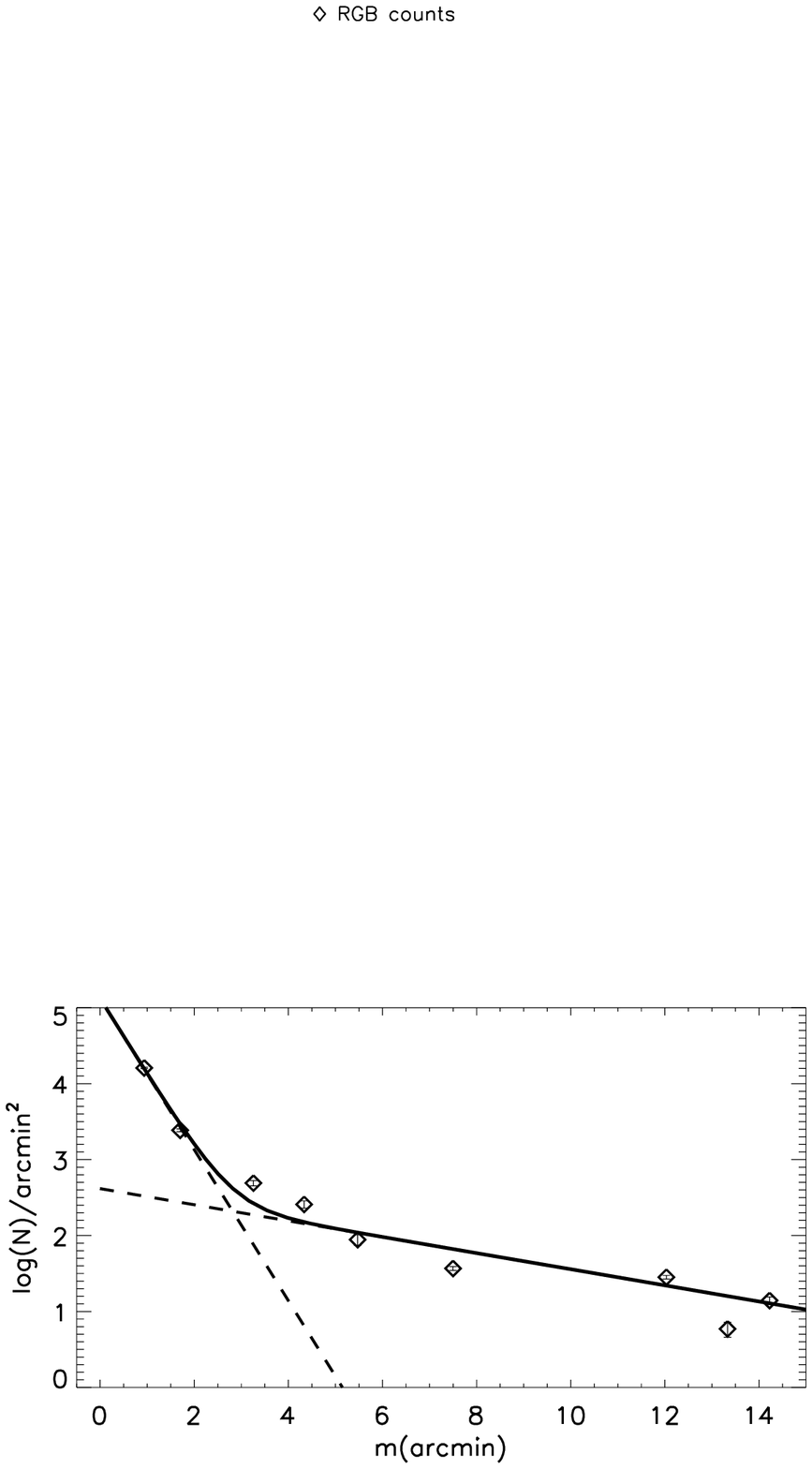}
\includegraphics[width=0.9\columnwidth]{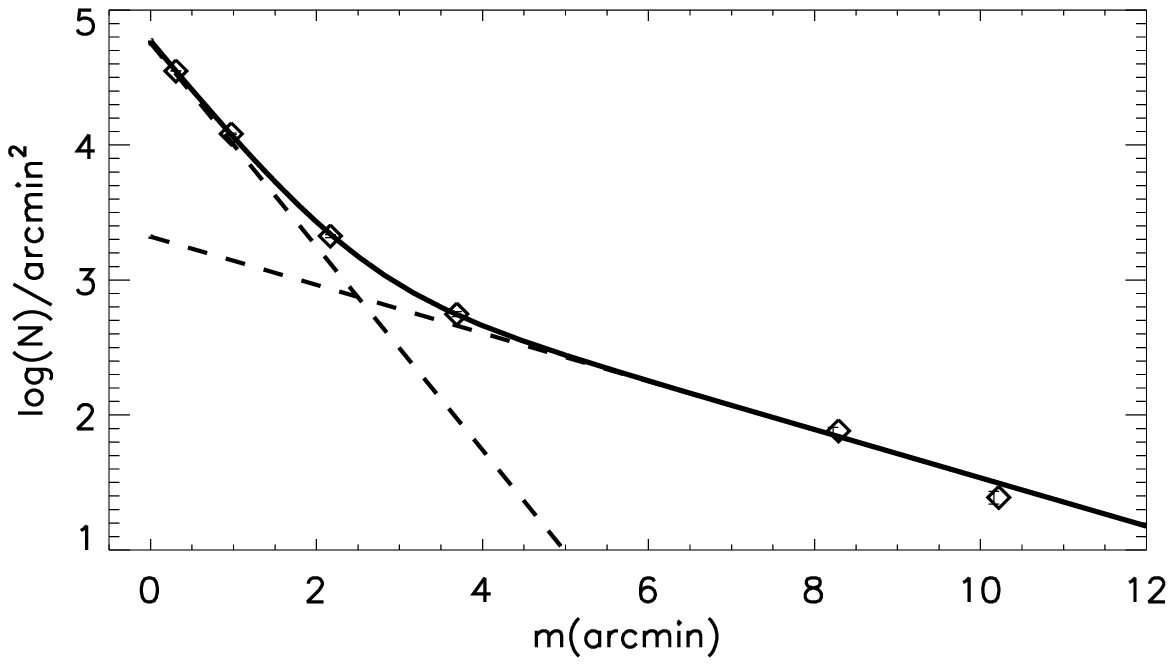}
   \caption[]{Double exponential fits (solid lines) to the data, with their single exponential components (dashed lines) indicated (NGC 1569 on the left and NGC 4449 on the right).}
\label{expfit}
\end{figure*}

While in principle we are able to fit a double exponential to the data (see Figure~\ref{expfit}), this does not mean that a thick disk is the favored explanation of the elevated density we see at large radii. Many of our fields are well away from the projected major axis, and in view of this a rounder component, i.e. a halo, seems more plausible.

\begin{figure*}[h]
  \centering
\includegraphics[width=0.9\columnwidth]{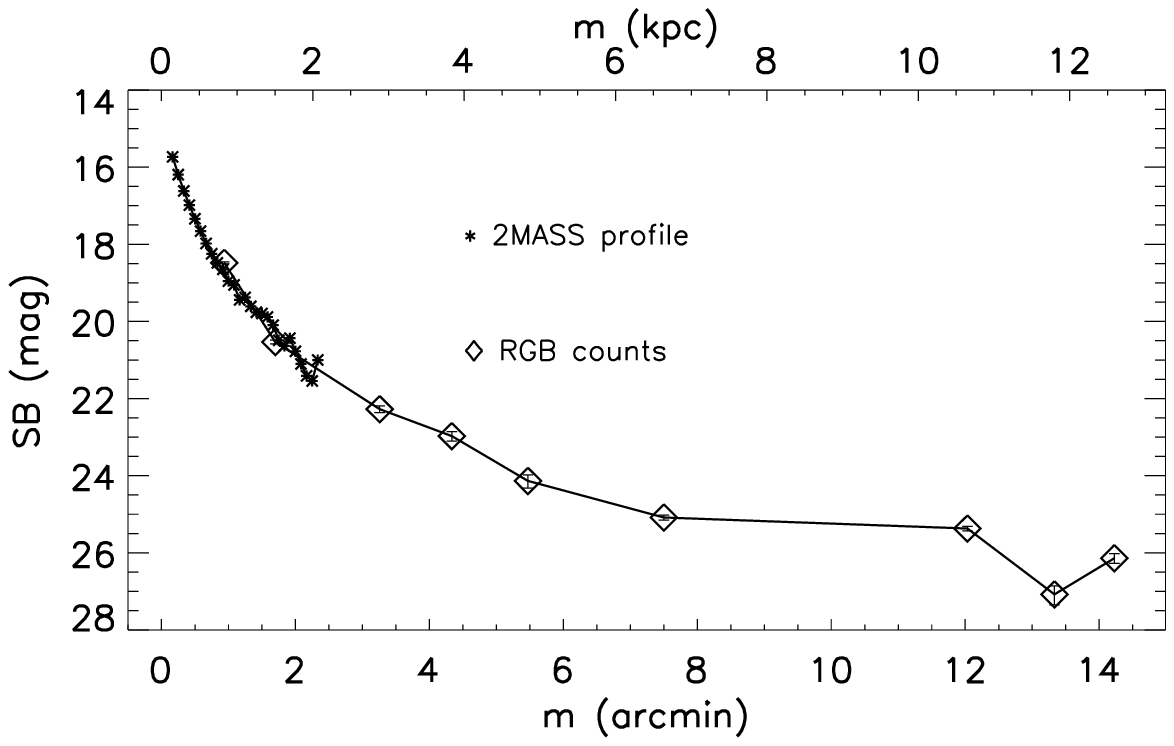}
\includegraphics[width=0.9\columnwidth]{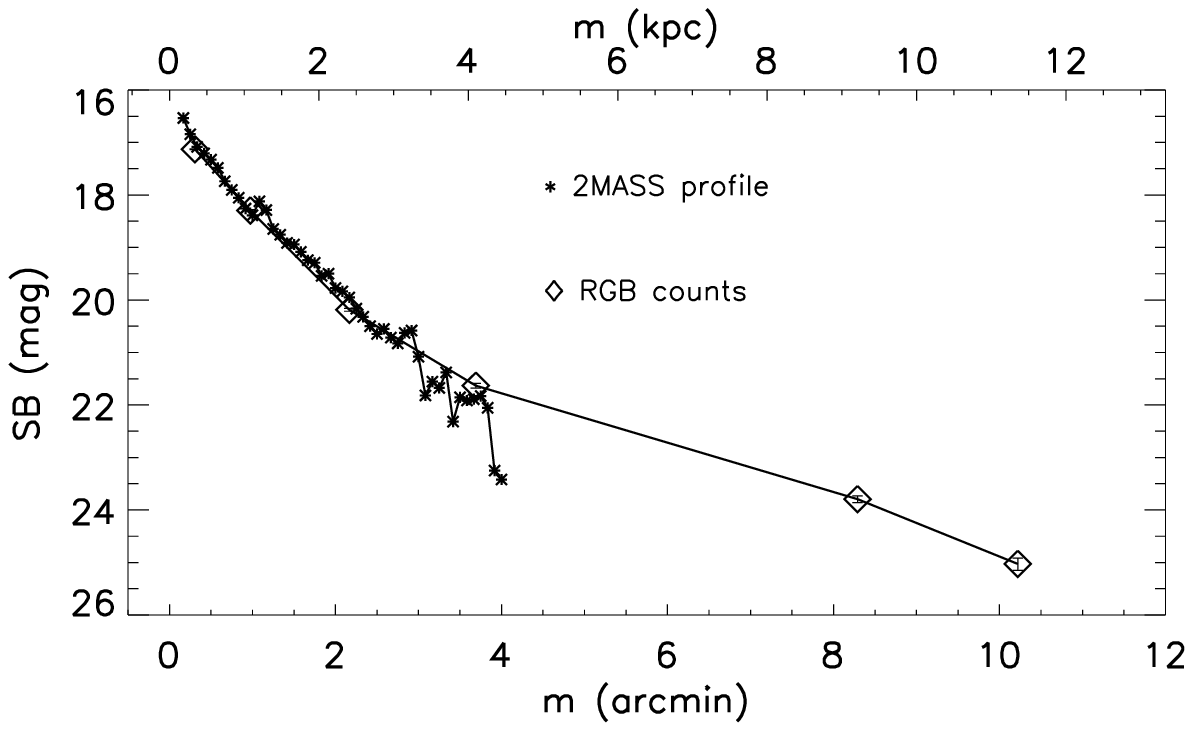}
\caption{Comparison of surface brightness profiles from 2MASS K-band
images with our RGB star counts for NGC 1569 (left) and NGC 4449
(right). A vertical shift was applied to the RGB profiles to line
the profiles up.}
\label{2mass}
\end{figure*}

\subsection{Integrated light profiles}
In order to perform a consistency check, we determined the profiles of
integrated light from \textit{2MASS K}-band images for comparison to
our RGB counts. $K$ is the reddest band in the \textit{2MASS} survey,
and is most dominated by light from RGB stars. Because intensity is
linearly proportional to the number of stars, one would expect the RGB
count profiles in Figure~\ref{density1} to match the measured
\textit{2MASS} surface brightness profiles after applying a vertical
shift.

The intensities at different major axis radii were calculated from the
2MASS images using the \textit{IRAF}\footnote{IRAF is distributed by
the National Optical Astronomy Observatory, which is operated by the
Association of Universities for Research in Astronomy (AURA) under
cooperative agreement with the National Science Foundation.} task
\textit{ellipse}, with the ellipticity, PA and axial ratio values
fixed to values typical at 2 arcmin radii (see Table~1). The
calibrated \textit{2MASS} magnitudes were then obtained by applying
the following formula: $mag = MAGZP - 2.5 \log_{10} (S)$, where the S
is the integrated, background-subtracted flux in ``$DN$'' (data
numbers) and $MAGZP$ is the zeropoint obtained from the image
header. We truncated the profiles at $\sim$0.4 \% of the sky value,
the same as the value adopted for profiles published in the
\textit{2MASS Large Galaxy Atlas} at
\textit{http://irsa.ipac.caltech.edu/}.

As can be seen in Figure~\ref{2mass}, the resulting profiles line up
continuously with the shifted RGB-count profiles, which was to be
expected since \textit{K}-band measures mostly light from RGB
stars. This also confirms that individual star counts are an excellent
way of following density profiles to large radii, where the integrated
light falls well below the sky background and becomes hard to
quantify. 

\section{Discussion and conclusions}

We obtained and analyzed HST/WFPC2 images of several fields in the
outer halos of the nearby starburst galaxies NGC 1569 and NGC
4449. The fields were imaged using the F606W and F814W filters. We
reduced the data and performed point source photometry to obtain CMDs
of $I$ vs. $V-I$. We used these CMDs to perform a study of stellar
population and density gradients in these galaxies. To maximize the
radial range available for analysis, we included in our study also the
CMDs (in the same bands) derived from our prior HST/ACS photometry
near the galaxy centers. Corrections for foreground contamination,
incompleteness, and differential dust extinction were calculated and
applied as necessary.

The HST/WFPC2 fields in our study are located roughly between 5 and 8
scale radii (exponential disk scale radius or effective radius) from
the galaxy centers. The corresponding elliptical (isophotal or major
axis) radii fall between 7 and 14 scale radii. Despite these
significant radii, we detect stars belonging to these galaxies in all
the fields. This implies that the galaxies have faint outer stellar
envelopes, and are not tidally truncated within the range of radii
addressed by our study. Similar results have been obtained for other star-forming dwarfs, see, e.g. the recent work by \cite{noel:07}, \cite{tikhonov:06b}, \cite{alonso-garcia:06}.

The CMDs of the HST/WFPC2 fields imply that the stars fall
predominantly in the evolutionary sequences of the RGB and the TP-AGB
(carbon star) phases. These phases correspond to stars of intermediate
and old ages (in excess of a few hundred Myr). Although younger stars
on the main sequence, blue loop, and red supergiant phases are present
in abundance near the centers of these starburst galaxies, they are
not seen in the outer HST/WFPC2 fields. This confirms results from our
own previous work on the central regions of these galaxies (\citealt{grocholski:08}, \citealt{annibali:08}), as well as
from studies of other galaxies in the literature (e.g. \citealt{tosi:01}, \citealt{tolstoy:09}), that the starburst
phenomenon tends to be strongly concentrated toward the galaxy
center. The presence of older stars at all radii indicates that the
starbursts occurred in pre-existing galaxies with prior star
formation, and were not the first star-formation episodes in an
otherwise primordial gas cloud.

We identified and quantified the TRGB magnitude in all HST/WFPC2
fields. The average WFPC2 TRGB magnitudes (not corrected for
foreground extinction) are $24.47 \pm 0.02$ (NGC 1569) and $24.06 \pm
0.07$ (NGC 4449). These values excellently agree with the
results obtained from our previously analyzed HST/ACS data: $24.47 \pm
0.04$ (NGC 1569; \citealt{grocholski:08}) and $24.02 \pm 0.05$ (NGC
4449; \citealt{annibali:08}), respectively. The new TRGB results
therefore support the galaxy distances derived in these previous studies.

To study the origin and characteristics of the intermediate-age/old
outer stellar envelopes of the sample galaxies (outside the central
starburst), we measured the observed radial profiles of the RGB color
and the number ratio of carbon to RGB stars. Gradients in these
quantities would be tell-tale signs of radial gradients in either the
age or metallicity of the population. However, to within the
uncertainties of our analysis, no gradients are detected (the 
maximum allowed gradients are -0.002 or +0.012 mag/kpc for NGC 1569, and -0.017 or +0.025 mag/kpc for NGC 4449). This does
not mean that no gradients exist, but merely that better data would be
needed to detect any gradients that might exist.

We measured the density profiles of RGB stars over the radial range
accessible to our study. At small radii we show that the profiles are
consistent with 2MASS measurements of near-IR integrated light. The
profiles were fitted with exponential, de Vaucouleurs, and power-law
models. While the latter profiles provide acceptable fits, exponential
profiles do not. Both galaxies have an excess of stars at large radii
above what a simple exponential model would predict. This indicates
that the intermediate-age/old stars at large radii at not merely the
outer part of an exponential disk. Instead, the observed density
profiles are more typical for an extended stellar halo. Such halos are
believed to be a common byproduct of hierarchical formation (but see 
also \citealt{stinson:09} for a discussion on how such halos can be formed 
in isolated galaxies, i.e. without external perturbations). Stellar halos 
have been found in all dwarf galaxies studied to a sufficient depth 
(see \citealt{stinson:09} and the extensive reference list therein). Our 
study shows that NGC 1569 and NGC 4449 have similar halos to those seen 
in other galaxies of similar morphology as well as that deep resolved 
stellar photometry enables one to distinguish between the profiles of 
the halos and the profiles of the star-forming disks they contain.

\begin{acknowledgements}

AR would like to thank Jennifer Mack for the help with the WFPC2 data reduction.
We thank Luca Angeretti, Laura Greggio, Enrico
V. Held, Claus Leitherer, Donatella Romano, Marco
Sirianni and Monica Tosi for their collaboration on other parts of the
HST projects GO-10585 and GO-10885.
We also want to thank the anonymous referee for his/her constructive comments that helped improve the manuscript.

Support for proposals GO-10585 and GO-10885 was provided by NASA
through grants from STScI, which is operated by AURA, Inc., under NASA
contract NAS 5-26555.

This publication makes use of data products from the Two Micron All
Sky Survey, which is a joint project of the University of
Massachusetts and the Infrared Processing and Analysis
Center/California Institute of Technology, funded by the National
Aeronautics and Space Administration and the National Science
Foundation.
\end{acknowledgements}

\bibliography{biblio3}

\end{document}